\documentclass[dvips]{article}

\usepackage{icrc2011}

\title{Propagation of Ultrahigh Energy Nuclei in the Magnetic Field of our Galaxy}

\newcommand{\etal}{\MakeLowercase{\textit{et al. }}} 
\shorttitle{G.~GIACINTI \etal UHE nuclei in the Galaxy}

\authors{G.~Giacinti$^{1}$, M.~Kachelrie\ss$^{2}$, D.~V.~Semikoz$^{1,3}$, G.~Sigl$^{4}$}
\afiliations{$^1$AstroParticle and Cosmology (APC), Paris, France\\ $^2$Institutt for fysikk, NTNU, Trondheim, Norway\\ $^3$Institute for Nuclear Research of the Russian Academy of Sciences, Moscow, Russia\\ $^4$II. Institut f\"ur Theoretische Physik, Universit\"at Hamburg, Germany}
\email{gwenael.giacinti@apc.univ-paris7.fr}

\abstract{In this work, we present detailed simulations for propagation of ultra-high energy (UHE) heavy nuclei, with $E \geq 60$\,EeV, within recent Galactic Magnetic Field (GMF) models~\cite{Giacinti:2010dk,Giacinti:2011uj}. We investigate the impacts of the regular and turbulent components of the GMF. We show that with UHE heavy nuclei, there is no one-to-one correspondence between the arrival directions of cosmic rays (CR) measured at Earth and the direction of their extragalactic sources. Sources can have several distorted images on the sky. We compute images of galaxy clusters and of the supergalactic plane in recent GMF models and show the challenges, and possibilities, of "UHECR astronomy" with heavy nuclei. Finally, we present a quantitative study of the impact of the GMF on the (de-)magnification of source fluxes, due to magnetic lensing effects. We find that for 60\,EeV iron nuclei, sources located in up to about one fifth of the sky would have their fluxes so strongly demagnified that they would not be detectable at Earth, even by the next generation of UHECR experiments.}

\begin{document}
\maketitle

\section{Introduction}

The Ultra-High Energy Cosmic Ray (UHECR) composition at the highest energies, above $E \sim 10^{19}$\,eV, is still a matter of debate. The measurements from the Pierre Auger Observatory (PAO) suggest a shift towards heavier nuclei~\cite{Collaboration:2010yv}. The muon data from the Yakutsk EAS Array were also found to be consistent with the PAO measurements~\cite{Glushkov:2007gd}. On the contrary, the measurements from HiRes~\cite{BelzICRC} as well as the preliminary results of Telescope Array~\cite{TA} are compatible with a proton composition.

The cosmic ray spectrum shape~\cite{hires-spec,auger-spec} is compatible with either a proton or an iron nuclei composition, for $E>10^{19}$\,eV. Heavy nuclei may be accelerated to higher energies than protons, which may lead to a transition towards heavy nuclei at the end of the spectrum. Additional information on the primary composition may be drawn from the difference of propagation between protons and heavy nuclei in the Galactic Magnetic Field (GMF). Until now, most studies of UHECR propagation in the GMF dealt with proton or light nuclei primaries, with the exception of References~\cite{Harari:1999it,Harari:2000he,Harari:2000az,Takami:2009qz,Vorobiov:2009km,Giacinti:2010dk,Giacinti:2011uj}.

In this work, we study the propagation of UHE iron nuclei with energies $E \geq 60$\,EeV in the GMF. Both the impacts of its regular -large scale- and turbulent -small scale- components are discussed. Quantities in given directions of the sky, such as the deflection angles on the celestial sphere of UHE nuclei between their sources and their arrival directions at Earth, are strongly model dependent. However, quantities averaged over the whole sky, such as the mean deflection angles, are considerably less model dependent for all tested recent GMF models. In this sense, we are able to present generic conclusions. We take here the Prouza and Smida (PS) model~\cite{PS,Kachelriess:2005qm} for the regular GMF. For the turbulent component, we construct it as a superposition of a finite number of plane waves, following Refs.~\cite{JG99,Giacinti:2011uj}.

Section~\ref{Images} deals with the image formation of UHE nuclei sources in the GMF, and discusses the possibilities and challenges of ``UHECR astronomy'' in case a heavy composition would be confirmed in the future. Section~\ref{Lensing} presents a study of (de-)magnification effects of individual UHE iron nuclei source fluxes. Such effects are due to magnetic lensing in the GMF, and can be considerable in a non-negligible fraction of the sky for heavy nuclei primaries.

\section{Images of UHE heavy nuclei sources}
\label{Images}

The shape of images of UHE proton \textit{point} sources on the celestial sphere has been discussed in detail in Refs.~\cite{Golup:2009cv,Giacinti:2009fy}. Except in directions to the Galactic plane and center, UHE protons are approximately deflected on the sky from their sources as $1/E$. At the highest energies, outside the Galactic plane, there is one-to-one correspondence between UHECR entering directions in the Galaxy and their directions on the sky as measured at Earth.

For heavy nuclei primaries, there is no one-to-one correspondence in large regions of the sky, even at the highest energies. In such regions, a given extragalactic source can have several images on the celestial sphere. In practice, nuclei from different images enter the Galaxy from the same direction, but at different positions in the physical space.


 \begin{figure}[!t]
  \vspace{5mm}
  \centering
  \includegraphics[width=0.43\textwidth]{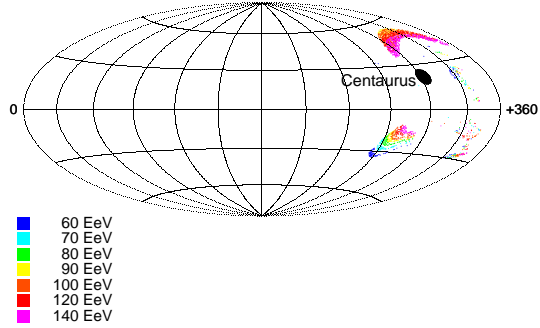}
  \caption{Image of the Centaurus galaxy cluster emitting UHE iron nuclei (from $E=60$ to 140\,EeV) deflected in the PS regular GMF model. Black disk for Centaurus position on the sky. Colours according to energies, see key.}
  \label{Figure4}
 \end{figure}

Before entering our Galaxy, UHECR may be deflected significantly in extragalactic magnetic fields (EGMF) which are expected to correlate with the Large Scale Structure (LSS) of galaxies but are otherwise poorly known. If UHECR deflections in EGMF are as low as in the Dolag \textit{et al.} scenario~\cite{Dolag:2003ra,Dolag:2004kp}, deflections of UHE iron nuclei would be negligible outside galaxy clusters. However, for a source located in a cluster, UHE nuclei would still be sufficiently deflected within the cluster to make it shine as a whole and make it appear as an extended source on the sky~\cite{virgoMF}. We compute in Fig.~\ref{Figure4} the image of the Centaurus galaxy cluster in the PS regular GMF model, for energies $E = 60 - 140$\,EeV, assuming the cluster contains one or several UHE iron nuclei source(s). The cluster is denoted by a black disk on the sky, and its cosmic ray arrival directions as measured at Earth are represented by small colour dots. If one also adds the contribution of the turbulent GMF, the UHECR arrival directions on the sky would be more spread. In some circumstances, its impact is less trivial than for UHE protons, especially at low Galactic latitudes~\cite{Giacinti:2011uj}. Due to the poor knowledge of the GMF~\cite{Jansson:2009ip}, Fig.~\ref{Figure4} should not be considered as a prediction. The exact image of a given source is strongly dependent on the GMF model. Even the two most recent benchmark models proposed in Ref.~\cite{Pshirkov:2011um} would result in different images. Nonetheless, the generic features visible in Fig.~\ref{Figure4} are quite typical for UHE heavy nuclei sources, in all Galactic magnetic field models. In most regions of the sky, heavy nuclei sources have several images. Some of them may be strongly distorted and (dis-)appear above given energy thresholds, such as the image at high latitudes in Fig.~\ref{Figure4}. The energy ordering of events on the sky may be very far from the $1/E$ ordering for protons, and in some cases, lower energy events may be closer to their sources than higher energy events. Images of more sources, and in other GMF models, are presented in Refs.~\cite{Giacinti:2010dk,Giacinti:2010ep}. As shown in Refs.~\cite{Giacinti:2010ep,ICRC_0171}, in some GMF models and for some sources, at least one of their images may still look like a proton image roughly enlarged by a factor $Z$, for nuclei of charge $Z$. In such cases, one may still be able to detect them with the algorithm presented in Refs.~\cite{Giacinti:2010ep,ICRC_0171}. However, in most cases, a better knowledge of the GMF than currently available is needed so as to find efficient algorithms of source detection.

 \begin{figure}[!t]
  \vspace{5mm}
  \centering
  \includegraphics[width=0.43\textwidth]{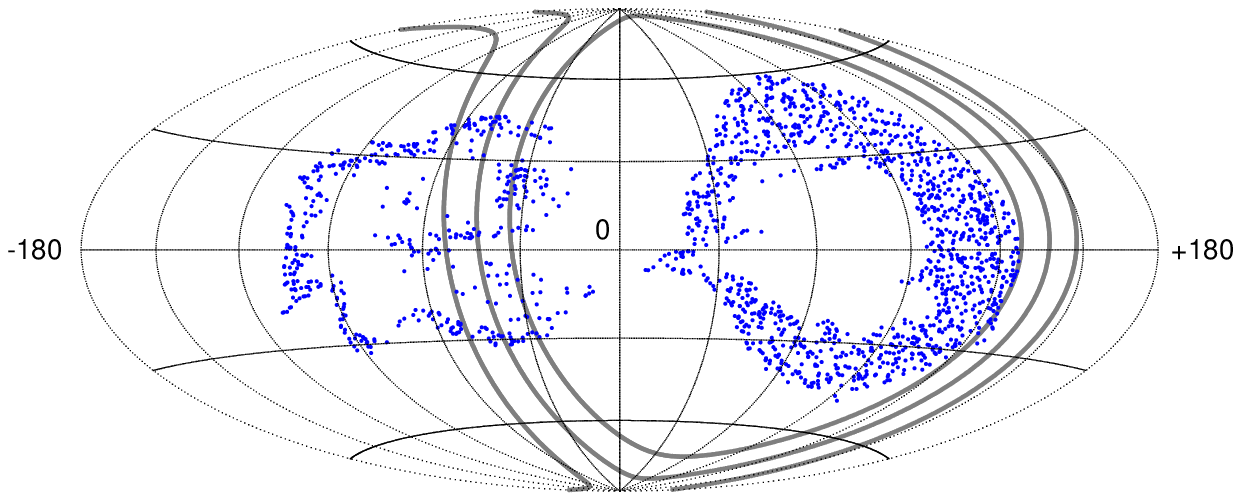}
  \includegraphics[width=0.43\textwidth]{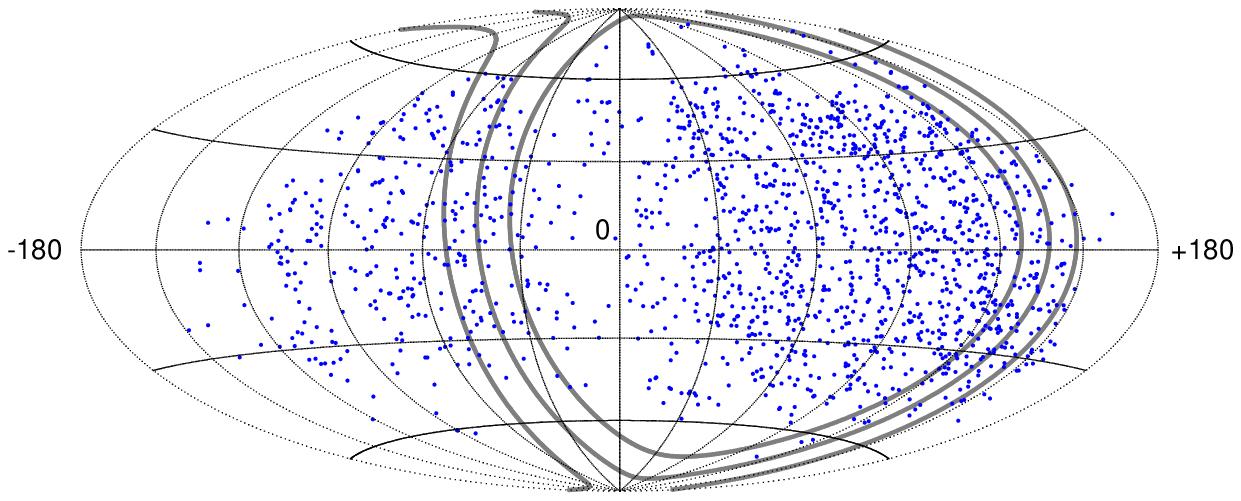}
  \caption{Images of the supergalactic plane, regarded as a $\pm10^{\circ}$ band in supergalactic latitude $b_{\rm SG}$, emitting 60\,EeV iron nuclei deflected in the GMF. PS model for the regular GMF. \textbf{Upper panel:} No turbulent field added; \textbf{Lower panel:} Additional turbulent component with $B_0=4\,\mu$G, $L_{c}=50$\,pc and $z_{0}=3$\,kpc. The blue points denote the CR arrival directions. The grey lines stand for the supergalactic plane and the $b_{\rm SG}=\pm10^{\circ}$ circles. Map centered on the Galactic center, with increasing $l$ from the left to the right.}
  \label{Figure56}
 \end{figure}

If UHECR deflections in EGMF are as large as in the Sigl \textit{et al.} scenario~\cite{Sigl:2004yk,Sigl:2004gi}, UHE heavy nuclei would still preferably be deflected within the LSS, with negligible deflections in the voids of the LSS. Therefore, even in this scenario, the UHECR directions when entering our Galaxy should still correlate with the LSS, even if they would not point back towards their sources -or clusters. As a first approximation, we study here the image of the supergalactic plane as a whole. Figure~\ref{Figure56} (upper panel) presents the 60\,EeV iron nuclei image of the supergalactic plane, in the PS regular GMF only. Dark blue dots correspond to the arrival directions of cosmic rays on the sky. We consider the supergalactic plane as a $\pm10^{\circ}$ strip in supergalactic latitude $b_{\rm SG}$. Grey lines stand for $b_{\rm SG}=-10^{\circ},0^{\circ},10^{\circ}$. Figure~\ref{Figure56} (lower panel) presents the same image when a non-zero turbulent component is added to the PS regular GMF model. We take here a turbulent field of correlation length $L_{c}=50$\,pc. We assume a Kolmogorov spectrum with variation scales between $L_{\min}=20$\,pc and $L_{\max}=200$\,pc, but for the chosen rigidities, our results do not depend on the magnetic field power spectrum. We take a spatial profile as in Ref.~\cite{Giacinti:2009fy} with a rms strength at Earth $B_0=4\,\mu$G and extension in the halo $z_{0}=3$\,kpc. For such parameters, the UHE iron nuclei arrival directions are significantly more spread on the sky, making it more difficult to recognize the initial supergalactic plane image without turbulent GMF. Since the turbulent component is still poorly known, the dependence of results on its parameters have been investigated in Ref.~\cite{Giacinti:2011uj}.

\section{Magnetic lensing in the GMF}
\label{Lensing}

Magnetic lensing of UHECR in the GMF results in the (de) magnification of source fluxes as seen at Earth. The flux detected at Earth for a given source is changed by a factor $\mathcal{A}$, compared to the flux one would have received if the GMF were set to zero. We call $\mathcal{A}$ the amplification factor. Depending on the direction on the sky, it may be either larger (magnification) or smaller (demagnification) than 1. For UHE proton sources, this effect is negligible in all GMF models ($\mathcal{A}\simeq1$), except for the unfavorable directions towards the Galactic plane and center. On the contrary, if UHECR are heavy nuclei, such effects would be substantial even at the highest energies $E \geq 60$\,EeV.

We compute $\mathcal{A}$ by backtracing a few times $10^5$ iron nuclei from the Earth to outside the Galaxy. The densities of outgoing nuclei are proportional to $\mathcal{A}$ -see References~\cite{Giacinti:2010dk,Giacinti:2011uj}.

\begin{figure}[!t]
  \vspace{5mm}
  \centering
  \includegraphics[width=0.43\textwidth]{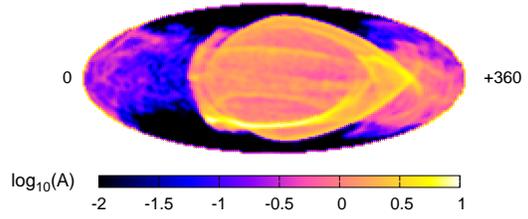}
  \caption{Logarithm of the amplification factor $\mathcal{A}$ of the flux of extragalactic 60\,EeV iron nuclei sources, depending on their position on the sky. PS model for the GMF. Black regions correspond to $\log_{10}(\mathcal{A})\leq-2$. Plot in Galactic coordinates, with the Galactic anti-center in the center.}
  \label{Figure1}
 \end{figure}

We present in Fig.~\ref{Figure1} the logarithm of the amplification factor, $\log_{10}(\mathcal{A})$, for 60\,EeV iron nuclei sources, depending on their positions on the sky. For the GMF, we take the PS model, with no turbulent component added ($B_0 = 0\,\mu$G). The black regions correspond to regions where $\log_{10}(\mathcal{A}) \leq -2$. Let us call them ``blind''. In such directions of the sky, fluxes of 60\,EeV iron nuclei sources would be demagnified more than 100 fold. Sources in blind regions cannot be detected neither by present nor near future experiments. In a few small regions (in white), the amplification factor is larger than 10. Only $\simeq55$\% of the sky has flux modifications by less than a factor $\simeq3$ ($-0.5<\log_{10}(\mathcal{A})<0.5$).

The exact amplification factor in a given direction of the sky is strongly model dependent. However, the distribution of regions with given amplification is less model dependent, for all recent GMF models.

 \begin{figure}[!t]
  \vspace{5mm}
  \centering
  \includegraphics[width=0.43\textwidth]{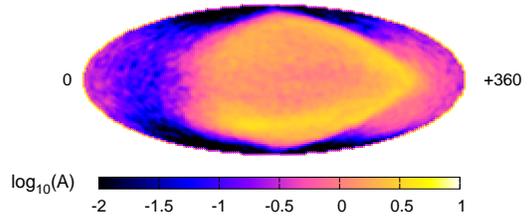}
  \caption{Same as in Fig.~\ref{Figure1}, with an additional turbulent component with $B_0=4\,\mu$G, $L_{c}=50$\,pc and $z_{0}=3$\,kpc.}
  \label{Figure2}
 \end{figure}

In Figure~\ref{Figure2}, we present the same plot as in Fig.~\ref{Figure1}, with a non-zero turbulent component added to the PS regular GMF. The turbulent field parameters that are considered in this figure are identical to those of Fig.~\ref{Figure56} (lower panel). One can see that the regions of extreme magnification or demagnification (in bright and dark colours) tend to shrink.

 \begin{figure}[!t]
  \vspace{5mm}
  \centering
  \includegraphics[width=0.43\textwidth]{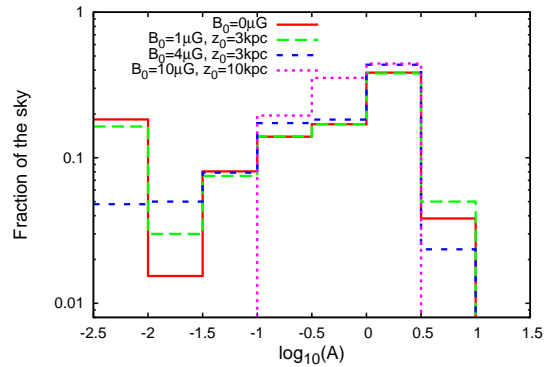}
  \caption{Fractions of the sky outside the Galaxy with amplifications $\mathcal{A}$. PS regular GMF model for all four histograms. Solid red line for no additional turbulent field. Green, blue and magenta lines for additional turbulent components with parameters respectively set to ($B_0=1\,\mu$G, $z_{0}=3$\,kpc), ($B_0=4\,\mu$G, $z_{0}=3$\,kpc) and ($B_0=10\,\mu$G, $z_{0}=10$\,kpc). Correlation length set to $L_{c}=50$\,pc. Bin of amplifications $\mathcal{A}$ below $10^{-2}$ for blind regions.}
  \label{Figure3}
 \end{figure}

We compute the extension on the sky of regions with given amplification factors, following the method of Ref.~\cite{Giacinti:2010dk}. Figure~\ref{Figure3} presents the results in bins of width $\Delta\log_{10}(\mathcal{A})=0.5$, both for the cases of Fig.~\ref{Figure1} (no turbulent field - red solid line) and Fig.~\ref{Figure2} (additional turbulent field with $B_0=4\,\mu$G and $z_{0}=3$\,kpc - blue thin dashed line). We also show the case of turbulent fields with ($B_0=1\,\mu$G, $z_{0}=3$\,kpc) and ($B_0=10\,\mu$G, $z_{0}=10$\,kpc), respectively with green thick dashed and magenta dotted lines. This plot confirms the above statement. For turbulent components that are stronger and more extended in the halo, regions of strong (de-) magnification globally shrink. For instance, the size of blind regions is divided here by $\sim3$ in the case of a turbulent field with $B_0=4\,\mu$G and $z_{0}=3$\,kpc. A strong and extended turbulent component would however not necessarily be helpful for the search of UHE heavy nuclei sources due to the larger blurring of their images -see Section~\ref{Images}. For the extreme case ($B_0=10\,\mu$G, $z_{0}=10$\,kpc), the fraction of the sky satisfying $-0.5<\log_{10}(\mathcal{A})<0.5$ reaches $\simeq80$\%. The maximum and minimum amplification factors also respectively decrease and increase when the turbulent field strength and spatial extension are increased.

\section{Conclusions and perspectives}
\label{Conclusion}

We have investigated here UHE heavy nuclei propagation in models of the Galactic magnetic field, taking into account both its regular and turbulent components. We have shown that several important effects appear in case of a heavy primary composition at the highest energies.

We computed in Section~\ref{Images} the images of an individual galaxy cluster and of the supergalactic plane, emitting UHE iron nuclei deflected in GMF models. We pointed out that for such a composition, non-trivial effects often appear. For instance, multiple and distorted images, or energy ordering of events far from the $1/E$ behaviour expected for proton sources, can be seen for sources located in large fractions of the sky, in all recent GMF models.

Except in a few favorable cases, a better knowledge of the GMF than currently available is still necessary in order to identify UHE heavy nuclei sources. Future radio telescopes, such as LOFAR and SKA, will enable one to improve the knowledge on the regular and turbulent GMF components, both in the Galactic disk and halo.

In Section~\ref{Lensing}, we studied the (de-)magnification of UHE iron source fluxes depending on their positions on the sky. This effect is due to magnetic lensing of UHECR in the GMF. While it is negligible at the highest energies for proton primaries, it becomes substantial for heavy nuclei. We showed that, at 60\,EeV, sources located in up to about one fifth of the sky can have their fluxes demagnified by more than a factor 100 at Earth, making them undetectable by present and next generation UHECR experiments. This fraction is reduced for stronger turbulent components. For a turbulent field strength of $4\,\mu$G at Earth and 3\,kpc extension in the halo, the fraction of invisible sky in the PS regular GMF model is divided by more than three. This would however not facilitate source detection and reconstruction because of the larger blurring on the sky of source images due to the turbulent component.

\section{Acknowledgments}
At Hamburg this work was supported by the DFG through the collaborative research center SFB 676.


\clearpage


\begin{thebibliography}{}

\bibitem{Giacinti:2010dk}
  G.~Giacinti, M.~Kachelrie\ss, D.~Semikoz, G.~Sigl,
  JCAP {\bf 1008}, 036 (2010)
  [arXiv:1006.5416 [astro-ph.HE]].

\bibitem{Giacinti:2011uj}
  G.~Giacinti, M.~Kachelrie\ss, D.~V.~Semikoz, G.~Sigl,
  Astropart.\ Phys.\  {\bf 35}, 192 (2011)
  [arXiv:1104.1141 [astro-ph.HE]].

\bibitem{Collaboration:2010yv}
  J.~Abraham {\it et al.} [PAO Collaboration],
  Phys.\ Rev.\ Lett.\  {\bf 104}, 091101 (2010)
  [arXiv:1002.0699 [astro-ph.HE]].

\bibitem{Glushkov:2007gd}
  A.~V.~Glushkov {\it et al.},
  JETP Lett.\  {\bf 87}, 190 (2008)
  [arXiv:0710.5508 [astro-ph]].

\bibitem{BelzICRC}
  J.~Belz {\it et al.}, for the HiRes Collaboration,
in {\em Proc. 31st ICRC, {\L}\'{o}d\'{z}, Poland, 2009\/}.

\bibitem{TA}
Y.~Tameda {\it et al.} [TA Collaboration], talk at the Japanese Physical Society meeting, March 26, 2010.

\bibitem{hires-spec}
  R.~U.~Abbasi {\it et al.},
  Astropart.\ Phys.\  {\bf 32}, 53 (2009)
  [arXiv:0904.4500 [astro-ph.HE]].

\bibitem{auger-spec}
  J.~Abraham {\it et al.}  [PAO Collaboration],
  Phys.\ Lett.\  B {\bf 685}, 239 (2010)
  [arXiv:1002.1975 [astro-ph.HE]].

\bibitem{Harari:1999it}
  D.~Harari, S.~Mollerach and E.~Roulet,
  JHEP {\bf 9908}, 022 (1999)
  [arXiv:astro-ph/9906309].

\bibitem{Harari:2000he}
  D.~Harari, S.~Mollerach and E.~Roulet,
  JHEP {\bf 0010}, 047 (2000)
  [arXiv:astro-ph/0005483].

\bibitem{Harari:2000az}
  D.~Harari, S.~Mollerach and E.~Roulet,
  JHEP {\bf 0002}, 035 (2000)
  [arXiv:astro-ph/0001084].

\bibitem{Takami:2009qz}
  H.~Takami and K.~Sato,
  Astrophys.\ J.\  {\bf 724}, 1456 (2010)
  [arXiv:0909.1532 [astro-ph.HE]].

\bibitem{Vorobiov:2009km}
  S.~Vorobiov {\it et al.},
  Nucl.\ Phys.\ Proc.\ Suppl.\  {\bf 196}, 203 (2009)
  [arXiv:0902.3123 [astro-ph.HE]].

\bibitem{PS}
  M.~Prouza and R.~Smida,
  Astron.\ Astrophys.\  {\bf 410} (2003) 1
  [arXiv:astro-ph/0307165].

\bibitem{Kachelriess:2005qm}
  M.~Kachelrie\ss, P.~D.~Serpico and M.~Teshima,
  Astropart.\ Phys.\  {\bf 26}, 378 (2006)
  [arXiv:astro-ph/0510444].

\bibitem{JG99}
  J.~Giacalone and J.~R.~Jokipii,
  ApJ {\bf 520}, 204 (1999).

\bibitem{Golup:2009cv}
  G.~Golup {\it et al.},
  Astropart.\ Phys.\  {\bf 32}, 269 (2009)
  [arXiv:0902.1742 [astro-ph.HE]].

\bibitem{Giacinti:2009fy}
  G.~Giacinti, X.~Derkx and D.~V.~Semikoz,
  JCAP {\bf 1003}, 022 (2010)
  [arXiv:0907.1035 [astro-ph.HE]].

\bibitem{Dolag:2003ra}
  K.~Dolag, D.~Grasso, V.~Springel and I.~Tkachev,
  JETP Lett.\  {\bf 79}, 583 (2004)
  [Pisma Zh.\ Eksp.\ Teor.\ Fiz.\  {\bf 79}, 719 (2004)]
  [arXiv:astro-ph/0310902].

\bibitem{Dolag:2004kp}
  K.~Dolag, D.~Grasso, V.~Springel and I.~Tkachev,
  JCAP {\bf 0501}, 009 (2005)
  [arXiv:astro-ph/0410419].

\bibitem{virgoMF}
  K.~Dolag, M.~Kachelrie\ss\ and D.~V.~Semikoz,
  JCAP {\bf 0901}, 033 (2009)
  [arXiv:0809.5055 [astro-ph]].

\bibitem{Jansson:2009ip}
  R.~Jansson, G.~Farrar, A.~Waelkens, T.~Ensslin,
  JCAP {\bf 0907}, 021 (2009)
  [arXiv:0905.2228 [astro-ph.GA]].

\bibitem{Pshirkov:2011um}
  M.~S.~Pshirkov, P.~G.~Tinyakov, P.~P.~Kronberg and K.~J.~Newton-McGee,
  Astrophys.\ J.\  {\bf 738}, 192 (2011)
  [arXiv:1103.0814 [astro-ph.GA]].

\bibitem{Giacinti:2010ep}
  G.~Giacinti, D.~V.~Semikoz,
  Phys.\ Rev.\  {\bf D83 } (2011)  083002.
  [arXiv:1011.6333 [astro-ph.HE]].

\bibitem{ICRC_0171}
  G.~Giacinti, D.~V.~Semikoz,
  ``Search for Nuclei Sources in the UHECR Data,''
  In {\em Proc. 32nd ICRC\/}.

\bibitem{Sigl:2004yk}
  G.~Sigl, F.~Miniati and T.~A.~Ensslin,
  Phys.\ Rev.\  D {\bf 70}, 043007 (2004)
  [arXiv:astro-ph/0401084].

\bibitem{Sigl:2004gi}
  G.~Sigl, F.~Miniati and T.~Ensslin,
  Nucl.\ Phys.\ Proc.\ Suppl.\  {\bf 136}, 224 (2004)
  [arXiv:astro-ph/0409098].

\end{thebibliography}
\end{document}